\documentclass[
	aps, prl, reprint,
	10pt, notitlepage, a4paper,
        floats, floatfix,
	amsmath, amssymb, amsfonts, %eqsecnum,
	superscriptaddress,
	showpacs, showkeys,
]{revtex4-1}

\usepackage{graphicx, epsfig, epstopdf} %include figure files
\usepackage{amsmath, amsfonts, amssymb}
\usepackage{bm, array}
\usepackage[usenames]{color}
\usepackage[
      colorlinks=true,
      linkcolor=blue,
      urlcolor=blue,
      filecolor=blue,
      citecolor=red,
      pdfstartview=FitV,
      pdftitle={},
      pdfauthor={},
      pdfsubject={},
      pdfkeywords={},
      pdfpagemode=None,
      bookmarksopen=true
]{hyperref}

\def\be{\begin{equation}}
\def\ee{\end{equation}}
\def\beq{\begin{eqnarray}}
\def\eeq{\end{eqnarray}}

\def\cR{{\cal R}}
\def\cS{{\cal S}}
\def\cT{{\cal T}}
\def\fg{{\mathfrak g}}

\renewcommand{\section}[1]{\paragraph{#1.
---}\phantomsection\addcontentsline{toc}{section}{#1}}

\begin{document}

\centerline{}
%\vskip 1cm
\title{Collapsing thin shells with rotation}

\author{T\'erence Delsate}
\email{terence.delsate@umons.ac.be}
\affiliation{Theoretical and Mathematical Physics Dpt.,  University of Mons, UMONS, 20, Place du Parc, B-7000 Mons, Belgium}

\author{Jorge V. Rocha}
\email{jorge.v.rocha@tecnico.ulisboa.pt}
\affiliation{CENTRA, Departamento de F\'{\i}sica, Instituto Superior T\'ecnico, Universidade de Lisboa, Avenida Rovisco Pais 1, 1049 Lisboa, Portugal}

\author{Raphael Santarelli}
\email{santarelli@ifsc.usp.br}
\affiliation{Instituto de F\'isica de S\~ao Carlos, Universidade de S\~ao Paulo, Caixa Postal 369, CEP 13560-970, S\~ao Carlos, S\~ao Paulo, Brazil}

\date{\today}

\begin{abstract}
We construct exact solutions describing the motion of rotating thin shells in a fully backreacted five-dimensional rotating black hole spacetime. The radial equation of motion follows from the Darmois-Israel junction conditions, where both interior and exterior geometries are taken to be equal angular momenta Myers-Perry solutions. We show that rotation generates anisotropic pressures and momentum along the shell. Gravitational collapse scenarios including rotation are analyzed and a new class of stationary solutions is introduced. Energy conditions for the matter shell are briefly discussed.
% We construct exact solutions describing the motion of rotating thin shells
%in a fully backreacted      %% (and also rotating)
% five-dimensional rotating black hole spacetime. The radial equation
% of motion follows from the Darmois-Israel junction conditions for matching inner
% and outer geometries, both of which are taken to be equal angular momenta
% Myers-Perry solutions.
% %% The restriction to equally spinning spacetimes amounts to a technical simplification.
% We show that the effect of rotation in the collapse is the appearance of
% anisotropic pressures and intrinsic momentum along the shell.
% %% We present two particularly interesting scenarios. The first
% %% one is the full collapse of a shell onto a pre-existing asymptotically flat
% %% black hole, starting at rest (in radial coordinate) from infinity. Secondly, we
% %% report on a stable stationary solution for a shell around a rotating black hole
% %% in asymptotically anti-de Sitter space.
% We discuss the case of full collapse and introduce a new class of stationary
% solutions. Energy conditions for the matter composing the thin shell are briefly 
% discussed. The weak energy condition is satisfied in the examples provided.
 \end{abstract}


\maketitle

%%%%%%%%%%%%%%%%%%%%%%%%%%%%%%%%%%%%%%%%%%%
%%%%%%%%%%%%%%%%%%%%%%%%%%%%%%%%%%%%%%%%%%%
\section{Introduction}
\label{sec:Intro}

Dynamical processes in rotating spacetimes are notoriously difficult to model analytically.
% There are by now many studies of non spherically symmetric
% collapse in numerical relativity, starting with~\cite{Nakamura:1981,
% Stark:1985da}. Among the subjects that have been driving the gravitational
% collapse literature are those of critical phenomena~\cite{Gundlach:2007gc} and
% cosmic censorship~\cite{Joshi:2012mk}, which greatly benefit from analytic
% results.
% However,
In fact, there exist very few exact solutions in closed form describing
collapsing matter carrying angular momentum. A few notable exceptions were
obtained in $2+1$ dimensions, where the problem naturally becomes more
tractable~\cite{Mann:2008rx, Vaz:2008uv}
(see~\cite{Nolan:2002zd,DiPrisco:2009zc} for four-dimensional, but cylindrically symmetric,
studies).
The aim of this Letter is to construct exact rotating solutions to the Einstein field equations in five spacetime dimensions involving matter shells around black holes. In particular, these solutions allow to address for the first time the effect of rotation on the gravitational collapse in an exact manner in a scenario with more than two spatial dimensions.

Our setup consists of a gravitating thin rotating shell in five dimensions, which backreacts on the geometry of a black hole. We allow for the presence of a cosmological constant. Our construction relies on the well known Darmois-Israel formalism~\cite{Israel:1966rt, Darmois} for matching two spacetimes along a spacelike hypersurface.
% The DI junction conditions relate the jump in first derivatives of the metric 
% across this interface with the stress-energy of matter along the hypersurface.
% In general, this shell will be moving in spacetime, either collapsing, bouncing
% or oscillating. In certain circumstances it is possible to find stationary shells
% around rotating black holes.
This framework was originally employed in~\cite{Peleg:1994wx} to study the spherical collapse of thin dust rings in $2+1$ dimensions. More recently it was extended to the rotating case and to shells with pressure in Ref.~\cite{Mann:2008rx} (see~\cite{Lindblom:1974bq} for an early study in four dimensions employing a slow rotation approximation).

The generalization to higher dimensions is not straightforward at all, the reason being that the spacetime will typically depend non trivially on polar angles in addition to a radial coordinate, thus rendering the matching procedure intractable on analytic grounds. However, it is possible to make progress by restricting to the class of higher (odd) dimensional black holes with all --- generically independent --- angular momenta equal: in this case the problem is cohomogeneity-1~\cite{Bizon:2005cp,Bizon:2006wk,Kunduri:2006qa}. For simplicity we develop the formalism in $4+1$ dimensions but similar results are expected for higher odd dimensions.

% This Letter is organized as follows. In the second section, we briefly review
% the 5D equally spinning black hole geometries. Then we discuss the
% matching procedure, which dictates a stress-energy tensor on the shell. 
% It turns out that the matter on the shell must take the form of an {\em imperfect} 
% fluid, with anisotropic pressures and intrinsic momentum. This is a novel 
% feature not present in the lower dimensional scenarios previously studied. 
% Next, we discuss energy conditions for the class of stress tensors implied by 
% the matching procedure. In the fifth section, by adopting a linear equation of state, 
% we derive a particularly simple equation of motion for the shell. The nature of the 
% solution depends upon the choice of various parameters. We do not perform an 
% exhaustive study of the parameter space, but as an illustrative application of the 
% formalism, we present two distinct solutions: the first describes the full collapse of 
% a rotating shell onto a pre-existing black hole in asymptotically flat space, and the 
% second corresponds to a stationary shell surrounding a rotating black hole.}

Throughout the manuscript we use geometrized units, by setting both the speed of light $c$ and Newton's constant $G$ equal to one.

 \smallskip
%%%%%%%%%%%%%%%%%%%%%%%%%%%%%%%%%%%%%%%%%%%
%%%%%%%%%%%%%%%%%%%%%%%%%%%%%%%%%%%%%%%%%%%
\section{Equally spinning black holes in five dimensions\label{sec:BHgeometry}}

The rotating black hole geometry generalizing the Kerr solution to higher dimensions is the well known Myers-Perry black hole~\cite{Myers:1986un}, which has been extended to include a cosmological constant in~\cite{Hawking:1998kw, Gibbons:2004js}.

Five-dimensional spacetimes admit two orthogonal rotation planes, and consequently two independent angular momenta. When the two angular momenta are equal, the isometry group of the solution gets enhanced and the metric functions can be written in terms of a single coordinate $r$~\cite{Kunduri:2006qa}:
\beq
  ds^2  &=&  - f(r)^2 dt^2 + g(r)^2 dr^2 + r^2 \widehat{g}_{ab} dx^a dx^b \nonumber\\
  && +\, h(r)^2 \left[ d\psi + A_a dx^a - \Omega(r) dt \right]^2\,,
\label{eq:metric}
\eeq
where
\beq
  g(r)^2  &=&  \left( 1 + \frac{r^2}{\ell^2} - \frac{2M\Xi}{r^2} + \frac{2Ma^2}{r^4} \right)^{-1}\,, 
\label{eq:metricfuncs1}\\
  h(r)^2  &=&  r^2 \left( 1 + \frac{2Ma^2}{r^4} \right)\,, \qquad \Omega(r) =  \frac{2Ma}{r^2 h(r)^2}\,,
\label{eq:metricfuncs2}\\
  f(r)  &=&  \frac{r}{g(r) h(r)}\,, \qquad \Xi = 1 - \frac{a^2}{\ell^2}\,,
\label{eq:metricfuncs3}
\eeq
%
%% (to include factors of $G$ we just need to replace $M \to GM$)  
and with $\widehat{g}_{ab}$ and $A=A_a dx^a$ given by
\be
  \widehat{g}_{ab} dx^a dx^b  =  \frac{1}{4} \left( d\theta^2 + \sin^2\theta \, d\phi^2  \right), \;\;
  A = \frac{1}{2} \cos\theta \, d\phi\,.
\ee

The line element in the form~\eqref{eq:metric} generalizes to higher odd
dimensions $D=2N+3$, with $N$ an integer~\cite{Kunduri:2006qa}. Here, we shall
restrict to the case $N=1$, for which the metric~\eqref{eq:metric} is a solution
of the vacuum Einstein equations with a cosmological constant, equal to
$\Lambda=-6\ell^{-2}$.
%%
%\be
%  R_{\mu\nu} = - 4\ell^{-2} g_{\mu\nu}\,,
%\ee
%%
%where $\ell$ is the AdS length.
The asymptotically flat case can be recovered by taking the limit $\ell\to\infty$.

The largest real root of $g^{-2}$ marks an event horizon which possesses the geometry of a homogeneously squashed $S^3$. The mass ${\cal M}$ and angular momentum ${\cal J}$ of the spacetime are given by~\cite{Kunduri:2006qa}
\be
  {\cal M} = \frac{\pi M}{4} \left( 3 + \frac{a^2}{\ell^2} \right)\,, \qquad
  {\cal J} = \pi M a \,.
\ee
%Keeping factors of $G$ the expressions would be
%%
%\be
%{\cal M} = \frac{\pi M}{4G} \left( 3 + \frac{a^2}{\ell^2} \right)\,, \qquad
%{\cal J} = \frac{\pi M a}{G} \,.
%\ee
%%

\smallskip
%%%%%%%%%%%%%%%%%%%%%%%%%%%%%%%%%%%%%%%%%%%
%%%%%%%%%%%%%%%%%%%%%%%%%%%%%%%%%%%%%%%%%%%
\section{Matching spacetimes\label{sec:implodingshell}}

We take two spacetimes with line element given by Eq.~\eqref{eq:metric} for the interior and exterior spacetime. This metric describes a family of black hole spacetimes, parametrised by the pair $(M,a)$. We further add an index $\alpha$ to these parameters since in general the inner and outer spacetimes have different parameters. We take $\alpha=+$ ($\alpha=-$) for the exterior (interior) geometry. Similarly, the metric functions in~(\ref{eq:metricfuncs1}-\ref{eq:metricfuncs3}) also acquire an index $\alpha$, reflecting the choice of mass and spin parameters.

Let $\Sigma=\{x^\mu : t=\cT(\tau),\, r =\cR(\tau) \}$ be the hypersurface along which the shell lies. This 4D surface can be parametrised by coordinates $y^i=\{\tau,\psi,\theta,\phi\}$. Denote by $\fg_{ij}^{(\alpha)}$ the induced metric on $\Sigma$ as determined from the exterior/interior solution, and by $k_{ij}^{(\alpha)}$ the associated extrinsic curvature, with $k^{(\alpha)}$ being its trace. The Darmois-Israel junction conditions can be expressed as
\begin{flalign}
  \fg_{ij}^{(+)} &= \fg_{ij}^{(-)} \equiv \fg_{ij}\,,
\label{eq:junction1}\\
  (k_{ij}^{(+)} &- k_{ij}^{(-)}) - \fg_{ij} (k^{(+)}-k^{(-)}) = - 8\pi G \cS_{ij} \,,
\label{eq:junction2}
\end{flalign}
where $\cS_{ij}$ stands for the surface stress-energy tensor.

Before computing the junction conditions it is useful~\cite{Mann:2008rx} to convert to the comoving frame to eliminate cross terms in the induced metrics, by making the following change of coordinates
\be
  d\psi \longrightarrow d\psi' + \Omega_\alpha(\cR(t)) dt\,.
\ee
The exterior and interior metrics then become
\beq
  ds_\alpha^2  &=&  - f_\alpha(r)^2 dt^2 + g_\alpha(r)^2 dr^2 + r^2 \widehat{g}_{ab} dx^a dx^b \\
  &+& h_\alpha(r)^2 \left[ d\psi' + A_a dx^a + (\Omega_\alpha(\cR(t))-\Omega_\alpha(r)) dt \right]^2.\nonumber
\label{eq:comoving}
\eeq
From now on we will drop the prime in $\psi'$ and assume we are in the comoving frame, in which case $\fg_{\tau j}=-\delta_{\tau j}$.

The first junction condition~\eqref{eq:junction1} implies that the combination
$Ma^2$ is invariant across the shell~\cite{footnote1}
% \footnote{Notice that in 3D no such restriction is obtained since the induced
% metrics on both sides of the shell trivially match.}
and also provides a
relation between $\left(\frac{d\cT}{d\tau}\right)^2$ and
$\left(\frac{d\cR}{d\tau}\right)^2$:
\begin{flalign}
  & M_+a_+^2 = M_-a_-^2\,, \label{eq:Ma2}\\
  & - f_\alpha(\cR)^2 \left(\frac{d\cT}{d\tau}\right)^2 + g_\alpha(\cR)^2 \left(\frac{d\cR}{d\tau}\right)^2 = -1\,.
\end{flalign}

The extrinsic curvature
% , from the perspective of the 5D embedding spacetime,
is obtained from $k_{\mu\nu}=(g_{\mu\sigma}-n_\mu n_\sigma) \nabla^\sigma n_\nu$,
where $n_\mu = f(r)g(r)\left(-\frac{d\cR}{d\tau} , \frac{d\cT}{d\tau} , 0, 0, 0 \right)$, is
the unit normal vector to the hypersurface $\Sigma$. In terms of the coordinates $y^i$ along the hypersurface, the induced extrinsic curvature is written as $k_{ij}=k_{\mu\nu} \frac{dx^\mu}{dy^i} \frac{dx^\nu}{dy^j}$.

The second junction condition~\eqref{eq:junction2} requires the shell stress-energy tensor to take the form of an
imperfect fluid with anisotropic pressure and intrinsic momentum:
\be
  \cS_{ij} = (\rho+P)u_i u_j + P\, \fg_{ij} + 2\varphi\, u_{(i} \xi_{j)} + \Delta P \, \cR^2 \widehat{g}_{ij}  \,,
\label{eq:stress}
\ee
where $u =\partial_\tau$ is the (normalized) fluid four-velocity, $\xi = h(\cR)^{-1} \partial_\psi$ and $\widehat{g}_{ij}dy^idy^j=\widehat{g}_{ab}dx^adx^b$. Collapse with fluids of this type was recently considered in~\cite{Herrera:2014wia}.
Observe that when $\Delta P=\varphi=0$ we retrieve the stress-energy tensor of a
perfect fluid~\cite{footnote2}.
% \footnote{For consistency of
% Eqs.~\eqref{eq:junction2} we must have
% ${\cal S}_{\tau\phi}=\frac{\cos\theta}{2}{\cal S}_{\tau\psi}$,
% ${\cal S}_{\psi\phi}=\frac{\cos\theta}{2} {\cal S}_{\psi\psi}$ and
% ${\cal S}_{\phi\phi}=\frac{\cos^2\theta}{4} {\cal S}_{\psi\psi} + \sin^2\theta \, {\cal S}_{\theta\theta}$.}.

We find that the junction conditions~\eqref{eq:junction2} are satisfied if and only if
\beq
  \rho &=& - \frac{(\beta_+-\beta_-)(\cR^2 h)'}{8 \pi \cR^3}\,, \;\; 
  \varphi = - \frac{({\cal J}_+-{\cal J}_-) (\cR h)'}{4 \pi^2 \cR^4 h}\,,
\label{eq:stress_components}\\
  P&=&\frac{h}{8\pi \cR^3} \left[\cR^2 (\beta_+-\beta_-)\right]', \;\;
  \Delta P=\frac{(\beta_+-\beta_-)}{8\pi}\left[\frac{h}{\cR}\right]', \nonumber
\eeq
%
%%Keeping factors of $G$ we would have:
%\beq
%  \rho &=& - \frac{(\beta_+-\beta_-)(\cR^2 h)'}{8 \pi G\, \cR^3}\,, \; 
%  \varphi = - \frac{({\cal J}_+-{\cal J}_-) (\cR h)'}{4 \pi^2 \cR^4 h}\,,
%\label{eq:stress_components}\\
%  P&=&\frac{h}{8\pi G\, \cR^3} \left[\cR^2 (\beta_+-\beta_-)\right]', \;
%  \Delta P=\frac{(\beta_+-\beta_-)}{8\pi G}\left[\frac{h}{\cR}\right]'. \nonumber
%\eeq
%%
%(Inserting everything into the second junction condition we learn that six out of ten equations are trivially satisfied, corresponding to components $\{\tau\theta\} , \{\tau\phi\}, \{\psi\theta\} , \{\psi\phi\} , \{\theta\phi\}  , \{\phi\phi\} $).
where we have introduced the quantities
\be
  \beta_\alpha \equiv f_\alpha(\cR) \sqrt{1+g_\alpha(\cR)^2 \left( \frac{d\cR}{d\tau} \right)^2}\,,
\label{eq:beta}
\ee
and where primes stand for $d/d\cR$. The constraint~\eqref{eq:Ma2} is already being used, so that $h_+(\cR)=h_-(\cR)\equiv h(\cR)$ and $h_+'(\cR)=h_-'(\cR)\equiv h'(\cR)$.
Note that the momentum $\varphi$ and the anisotropic pressure term $\Delta P$
necessarily vanish in the limit of zero rotation. 
We stress that the component $\varphi$ controls the difference between the angular momentum of the outer and inner spacetimes.
% These results strongly suggest that in the case of collapse with rotation,
% even when the amount of symmetries is lower than in this study,
% some momentum and anisotropic pressure should be dynamically generated.

The strategy we have adopted, relying on the Darmois-Israel matching formalism,
improves on the perturbative approach developed in Ref.~\cite{Rocha:2014gza}, in
the sense that the solutions constructed herein account for all backreaction
effects. In~\cite{Rocha:2014gza} it was found that the shell was required to be
corotating with the spacetime~\cite{footnote3}
% \footnote{A similar restriction is not obtained in three
% dimensions~\cite{Rocha:2011wp} due to the triviality of  the BTZ black hole
% geometry~\cite{Banados:1992wn}.} 
but this can now be relaxed at the expense of
the fluid acquiring intrinsic momentum and anisotropic pressure.

\smallskip
%%%%%%%%%%%%%%%%%%%%%%%%%%%%%%%%%%%%%%%%%%%
%%%%%%%%%%%%%%%%%%%%%%%%%%%%%%%%%%%%%%%%%%%
\section{Energy conditions}

Energy conditions for imperfect fluids, such as~\eqref{eq:stress}, have been
studied in~\cite{Kolassis:1988} and are most easily formulated in terms of the
eigenvalues of the stress-energy tensor, say $\lambda_0,\ \lambda_i,
\;\; (i=1,2,3)$, where $\lambda_0$ is the eigenvalue associated with the
time-like eigenvector. The weak energy condition (WEC) reads~\cite{Wald:1984rg}
\be
  -\lambda_0\geq 0\,,\qquad \lambda_i - \lambda_0\geq0\,,
\ee
while the less restrictive null energy condition simply omits the first inequality.

In the specific case of the fluid~\eqref{eq:stress}, the eigenvalues are given by
\beq
  &&\lambda_0 = \frac{P-\rho}{2} -\sqrt{\left(\frac{P+\rho}{2}\right)^2-\varphi^2}\,,\nonumber\\
  &&\lambda_1 =\frac{P-\rho}{2} +\sqrt{\left(\frac{P+\rho}{2}\right)^2-\varphi^2}\,, \\
  &&\lambda_2=\lambda_3= P+\Delta P\,.\nonumber
\eeq
 This must be supplemented with the constraint $\rho+P\geq0$,  ensuring that
 the eigenvector corresponding to $\lambda_0$ is time-like~\cite{footnote4}.
%  \footnote{Otherwise, the role of $\lambda_0$ and $\lambda_1$ are interchanged
% and the WEC cannot be fulfilled.}.
% is the eigenvalue corresponding to the time-like eigenvector 
% of the stress-energy tensor.

\smallskip
%%%%%%%%%%%%%%%%%%%%%%%%%%%%%%%%%%%%%%%%%%%
%%%%%%%%%%%%%%%%%%%%%%%%%%%%%%%%%%%%%%%%%%%
\section{Shell equation of motion}

Let us assume a linear equation of state (EoS) of the form $P = w \rho$. Using
Eqs.~\eqref{eq:stress_components} this EoS translates into
\be
  \frac{[\cR^2 (\beta_+ - \beta_-)]'}{\cR^2 (\beta_+ - \beta_-)} = -w \frac{[\cR^2 h]'}{\cR^2 h}\,,
\ee
which admits the general solution
\be
  \beta_+-\beta_- = - \frac{m_0^{1+3w/2}}{\cR^{2(1+w)} h(\cR)^w}\,.
\ee
Here, $m_0$ is a positive constant with dimensions of mass. For the case of a non-rotating shell of dust ($a,w=0$) it corresponds to the initial rest mass of the shell.

Employing expression~\eqref{eq:beta}, this solution provides an equation for the radial motion of the shell,
\be
  \dot \cR^2 + V_{\rm eff}(\cR) =0\,,
\label{eq:motion}
\ee
where $\dot{\cR}\equiv d\cR/d\tau$ and the effective potential $V_{\rm eff}$ is given by
\beq
  && V_{\rm eff}(\cR) = 1 + \frac{\cR^2}{\ell^2} + \frac{2Ma^2}{\cR^4} + \frac{2Ma^2}{\ell^2\cR^2} - \frac{M_++M_-}{\cR^2} \nonumber\\
  && \qquad - \left(\frac{M_+-M_-}{m_0}\right)^2 \left(\frac{\cR^2}{m_0}\right)^{3w} \left(1+\frac{2Ma^2}{\cR^4}\right)^{w-1}\nonumber\\
  && \qquad -\frac{1}{4} \left(\frac{m_0}{\cR^2}\right)^{2+3w} \left(1+\frac{2Ma^2}{\cR^4}\right)^{1-w}.
\label{eq:potential}
\eeq

Given a set of parameters, namely the inverse curvature scale $\ell$ (including the flat limit $\ell\rightarrow\infty$), the interior and exterior mass parameters, $M_+,\ M_-$, the value of $Ma^2$ and the fluid parameters $w$ and $m_0$, a solution describing the collapse can be found by solving~\eqref{eq:motion}.
% If desired, it is possible to eliminate the velocity using Eq.~\eqref{eq:motion},
% and the acceleration using $\ddot R = - \frac{1}{2}V_{\rm eff}'(\cR)$.

First, let us examine the asymptotic behavior of the potential. For large values of $\cR$, it reads
\be
  V_{\rm eff} \approx 1+\frac{\cR^2}{\ell^2}
  - \left(\frac{\Delta M}{m_0}\right)^2 \left(\frac{\cR^2}{m_0}\right)^{3w}
  - \frac{1}{4} \left(\frac{m_0}{\cR^2}\right)^{2+3w},
\ee
while for small radii it reduces to
\beq
  V_{\rm eff} &\approx& \frac{2 Ma^2}{\cR^4} - \frac{M_++M_-}{\cR^2}
  - \frac{1}{4} \left(\frac{2Ma^2}{m_0^2}\right)^{1-w} \!\! \left(\frac{m_0}{\cR^2}\right)^{4+w}\nonumber\\
  &&- \left(\frac{2Ma^2}{m_0^2}\right)^{w-1} \left(\frac{\Delta M}{m_0}\right)^2 \left(\frac{\cR^2}{m_0}\right)^{2+w},
\eeq
where we have defined $\Delta M \equiv M_+-M_-$. (We have kept the subleading terms that dominate in the flat or nonrotating limits.) For $w=0$, we recover the AdS barrier at large radius (for positive $\ell^{-2}$) and the centrifugal repulsion term $\sim \cR^{-4}$ at small radius. However, observe that the confining potential is absent if $w>1/3$ or $w<-1$, and the centrifugal repulsion is counteracted by a term $\sim \cR^{-8-2w}$ when $w>-2$.
We also note that in the flat limit, the only possibility to describe a collapse starting at rest from infinity occurs for the choice of dust ($w=0$) and only if the condition $\Delta M = m_0$ is satisfied. For a such a collapse, the energy of the spacetime will be enhanced by the rest mass of the shell.
% 
% Finally, we present here the asymptotic behavior of the
% stress-energy tensor components in the AdS case:
% %
% \beq
%   &&\rho\approx\frac{3  m_0^{\frac{3 w}{2}+1} \cR^{-3 (w+1)}}{8\pi }\,,
%   \qquad
%   P= w \rho,\\
%   &&\Delta P \approx\frac{Ma^2 m_0^{\frac{3 w}{2}+1} \cR^{-3 w-7}}{2 \pi }
%   \,,\qquad
%   \varphi \approx -\frac{\Delta \cal J}{2 \pi^2 \cR^4}\,.
% \eeq
% %
% In the flat limit, the leading order acquires a non trivial dependence on $w$ and is straightforward to compute.

\smallskip
%%%%%%%%%%%%%%%%%%%%%%%%%%%%%%%%%%%%%%%%%%%
%%%%%%%%%%%%%%%%%%%%%%%%%%%%%%%%%%%%%%%%%%%
\section{Full collapse onto a black hole}

As an illustration, we now provide an explicit example of a rotating dust thin shell collapsing from rest at infinity onto a (small) rotating black hole in an asymptotically Minkowski spacetime. This is to the best of our knowledge the first exact solution of this sort in more than three spacetime dimensions.

\begin{figure}[t]
\begin{center}
 \includegraphics[width=8.6cm]{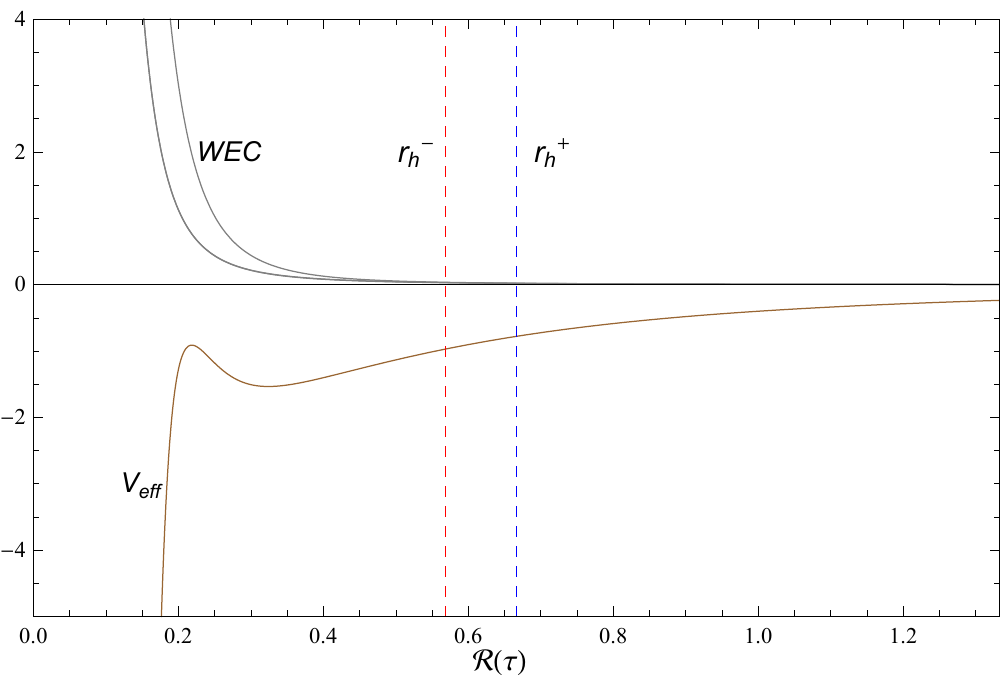}
\end{center}
\vspace{-0.5cm}
\caption{Gravitational collapse of a rotating shell (with $w=0$) in five-dimensional
asymptotically flat spacetime. The functions $-\lambda_0,\ \lambda_1-\lambda_0,\ \lambda_2-\lambda_0$
are collectively denoted by WEC. The curves for $-\lambda_0$ and $\lambda_2-\lambda_0$ nearly coincide for the solution displayed.
The horizon radius of the inner (outer)
geometry is denoted by $r_h^-$ ($r_h^+$). Note that the WEC
are satisfied throughout the collapse, all the way up to the singularity.
This particular plot corresponds to parameters
$M_-=0.2,\ M_+=0.25,\ Ma^2=0.01236,\ m_0=\Delta M= 0.05$ (in arbitrary units), which are chosen in such a way that the
collapse is complete.} 
 \label{fig:flatcollapse}
\end{figure}

In accordance with the above results we now take $\ell\to\infty$, set $w=0$ and $m_0=\Delta M$. It is easy to find conditions on the remaining parameters $\{Ma^2, M_+, M_-\}$ so that the effective potential is strictly negative (corresponding to a full collapse) and the eigenvalues of the stress-energy tensor are real and nonnegative (required for the weak energy conditions to be satisfied). In Figure~\ref{fig:flatcollapse} we show these quantities as functions of the radial location of the shell. A convenient choice of parameters was made for illustration purposes but we stress that no fine tuning is necessary.

This dynamical solution should be thought of as follows. Initially, the shell is located at $\cR\to\infty$ with vanishing radial velocity. It then collapses according to Eq.~\eqref{eq:motion}. The spacetime in the range $r<\cR(\tau)$ is described by the inner geometry and consists of a rotating black hole with a horizon at $r=r_h^-$ and spin parameter $a_- = \sqrt{Ma^2/M_-}$. At larger radii, the spacetime is the exterior of a black hole with horizon radius $r_h^+$ and spin parameter $a_+=\sqrt{Ma^2/M_+}$. As the shell crosses the radius $r_h^+$, a new apparent horizon forms at that location. Both the mass and spin of the initial black hole have changed. Eventually the shell falls into the curvature singularity at $r=0$, where the stress-energy components also blow up.

Note that none of the collapsing solutions we have described violate the cosmic censorship conjecture~\cite{Penrose:1969pc}: the identity~\eqref{eq:Ma2} implies that if the interior solution is sub-extremal the global solution after full collapse is similarly sub-extremal.
%(Positivity of $m_0$ requires $M_+\geq M_-$, which together with Eq.~\eqref{eq:Ma2} implies that $a_+\leq a_-$.)

\smallskip
%%%%%%%%%%%%%%%%%%%%%%%%%%%%%%%%%%%%%%%%%%%
%%%%%%%%%%%%%%%%%%%%%%%%%%%%%%%%%%%%%%%%%%%
\section{Stationary solution in AdS}

Given the `confining' nature of the effective potential in the presence of a negative cosmological constant (and when $-1<w<1/3$), as well as the existence of a centrifugal barrier, we expect that stationary solutions consisting of a rotating shell hovering around a black hole exist in AdS.

\begin{figure}[t]
\begin{center}
\vspace{-0.2cm}
 \includegraphics[width=8.72cm]{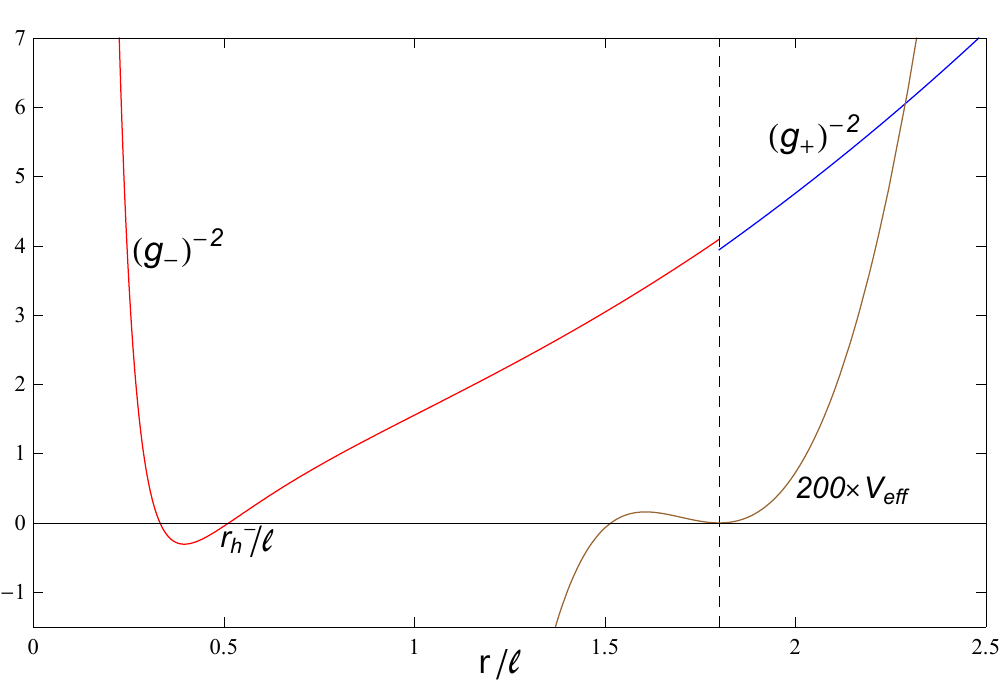}
\end{center}
\vspace{-0.462cm}
\caption{Stable configuration of a stationary rotating shell in AdS. The dashed
line represents the location $\cR_*$ of the shell. We plot the metric function
$g(r)^{-2}$, whose roots mark horizons. In the domain $r>\cR$ this function is
given by $g_+(r)^{-2}$ (blue) and for $r<\cR$ by $g_-(r)^{-2}$ (red). Notice the
discontinuity at $r=\cR_*$. The event horizon radius $r_h^{-}$ is given by the
largest real root of ${g_-}^{-2}$. The brown curve represents the effective
radial potential (amplified to ease interpretation) and $r=\cR_*$ corresponds to
a local minimum of the potential.
We chose the following values of parameters for this plot:
$m_0/\ell^2 = 0.324,\ w = 0.285,\ Ma^2/\ell^4 = 0.02,\ \cR _*/\ell = 1.8$.} 
 \label{fig:stationaryshell}
\end{figure}

Such solutions can be constructed as follows. We assume that (i) the radial location of the shell is constant, i.e., $\cR(\tau)=\cR_*$, implying that $V_{\rm eff}(\cR_*) = 0$, and that (ii) it corresponds to a local minimum of the potential,
$V'_{\rm eff}(\cR_*)=0$ and $V''_{\rm eff}(\cR_*)>0$. In general, these
conditions involve high order polynomial equations in $\cR_*$. On the other
hand, the conditions are simple quadratic equations in $M_+$ and $M_-$ (see
Eq.~\eqref{eq:potential}).
%
% 
% %
% \textcolor{blue}{
% \beq
%   &&M_+ + M_- = \frac{\cR^2}{(1+3 w)+(3+w)\frac{2 Ma^2}{\cR^4}} \left(1+\frac{2 Ma^2}{\cR^4}\right)\nonumber\\
%   && \!\!\!\!\!\!\!\!\!\!\!\! \times \left[ \frac{\cR^2}{\ell^2} \left(-1+3 w+(3+w)\frac{2 Ma^2}{\cR^4}\right) + \left(3w+(4+w)\frac{2 Ma^2}{\cR^4}\right) \right. \nonumber\\
%   && \!\!\!\!\!\!\!\!\!\! \left. -\frac{1}{2}\left(\frac{m_0}{\cR^2}\right)^{2+3w} \left(1+\frac{2 Ma^2}{\cR^4}\right)^{-w}\left(1+3w+(3+w)\frac{2 Ma^2 }{\cR^4}\right) \right] , \nonumber\\
% %
% \\
% %
%   &&M_+ - M_- = \frac{m_0}{2}\left(\frac{m_0}{\cR^2}\right)^{\frac{3}{2}w}\left(1+\frac{2 Ma^2}{\cR^4}\right)^{1-\frac{w}{2}} \nonumber\\
%   && \;\; \times \left[\left(8\frac{\cR^2}{L^2}+4-\frac{8 Ma^2}{\cR^4}\right)\left(1+3w+(3+w)\frac{2 Ma^2}{\cR^4}\right)^{-1} \right. \nonumber\\
%   && \qquad \left. +\left(\frac{\text{m0}}{\cR^2}\right)^{2+3 w} \left(1+\frac{2 Ma^2}{\cR^4}\right)^{-w} \right]^{\frac{1}{2}} .
% \eeq
% }
%
Therefore, it is more convenient to fix the masses of the inner and outer space in terms of the remaining parameters, $\{w, m_0, \cR_*, Ma^2\}$. This can be done in such a way to ensure that $\cR=\cR_*$ is a stable point. Then one just needs to scan the parameter space to obtain solutions for which the inner geometry possesses an event horizon, while the outer geometry has no horizons, and satisfy the WEC. It is easy to find such solutions and we show an example in Figure~\ref{fig:stationaryshell}.

We mention in passing the existence of unstable stationary solutions in
flat space, where the gravitational attraction is balanced by the centrifugal repulsion.
These configurations are obtained by tuning the parameters so that the potential of
Fig.~\ref{fig:flatcollapse} acquires a maximum with $0$ value. 
% We also remark that asymptotically flat stable 
% configurations, with $V_{\rm eff}(\cR)=0=V_{\rm eff}'(\cR)$, also exist in the
% following very particular cases:
% (i) when the outer geometry is extremal, the shell sits at the
% location of this degenerate horizon and the inner geometry describes a naked
% singularity (ii) when the inner geometry is extremal, in which case the shell is
% located inside the outer horizon.

\smallskip
%%%%%%%%%%%%%%%%%%%%%%%%%%%%%%%%%%%%%%%%%%%
%%%%%%%%%%%%%%%%%%%%%%%%%%%%%%%%%%%%%%%%%%%
\section{Discussion\label{sec:conc}}

In this Letter we showed it is possible to address non spherical gravitational collapse with analytic tools and nonperturbatively, even without restricting to lower dimensional spacetimes. In particular, we developed the formalism to study the motion of rotating thin shells in equal angular momenta black hole spacetimes in five dimensions, allowing for a cosmological constant.

The consideration of five dimensions is not physically fundamental: we regard it as a technical crutch to make progress analytically. We also stress that, while the symmetry required for the success of our analysis is enhanced relative to the case with arbitrary angular momenta ($U(2)$ compered to $U(1)^2$), it is still more generic than the spherically symmetric situation, which has isometry group $SO(4)\simeq SU(2)\times SU(2)$.

The most important lesson borne out by this work is that anisotropic pressures and intrinsic momentum are necessarily generated when rotation is present. This result is independent of the shell equation of state and is expected to hold even in less symmetric collapse scenarios. We further believe that the departure of the stress-energy tensor from the perfect fluid form is typical of collapse with rotation also in less idealized cases, in which the shell has finite thickness. In particular, we expect numerical simulations of gravitational collapse to display this feature.
%As a consequence, special care must be devoted to analyzing the energy conditions when rotation is present.

We presented two examples of exact solutions: a full collapse with rotation onto an asymptotically flat black hole and a stationary configuration of matter around a rotating black hole in AdS.  
% Solutions belonging to the first class cannot lead to cosmic censorship violation, 
% assuming the shell starts at rest from infinity. Solutions within the second class 
% are stable against perturbations of the radius of the shell.
Our analysis of the parameter space of solutions is far from being exhaustive. Such an extension is planned to be studied elsewhere.

\smallskip
%%%%%%%%%%%%%%%%%%%%%%%%%%%%%%%%%%%%%%%%%%%
%%%%%%%%%%%%%%%%%%%%%%%%%%%%%%%%%%%%%%%%%%%
\section{Acknowledgements}

We thank Vitor Cardoso, Don Marolf and Paolo Pani for useful discussions and
comments, and Vincenzo Vitagliano for bringing Ref.~\cite{Kolassis:1988} to our
attention.
J.~V.~R. would like to thank the University of California Santa Barbara for hospitality during completion of a portion of this work.
J.~V.~R. is supported by {\it Funda\c{c}\~ao para a Ci\^encia e Tecnologia} (FCT)-Portugal through Contract No.~SFRH/BPD/47332/2008. R.~S. is supported by Grant No. \#2012/20039-6, S\~ao Paulo Research Foundation (FAPESP).

%%%%%%%%%%%%%%%%%%%%%%%%%%%%%%%%%%%%%%%%%%%
%%%%%%%%%%%%%%%%%%%%%%%%%%%%%%%%%%%%%%%%%%%


\begin{thebibliography}{99}
\bibitem{footnote1}
  Notice that in 3D no such restriction is obtained since the induced
  metrics on both sides of the shell trivially match.
\bibitem{footnote2}
For consistency of
Eqs.~\eqref{eq:junction2} we must have
${\cal S}_{\tau\phi}=\frac{\cos\theta}{2}{\cal S}_{\tau\psi}$,
${\cal S}_{\psi\phi}=\frac{\cos\theta}{2} {\cal S}_{\psi\psi}$ and
${\cal S}_{\phi\phi}=\frac{\cos^2\theta}{4} {\cal S}_{\psi\psi} + \sin^2\theta
\, {\cal S}_{\theta\theta}$.
\bibitem{footnote4}
  Otherwise, the role of $\lambda_0$ and $\lambda_1$ are interchanged
and the WEC cannot be fulfilled.
\end{thebibliography}
\end{document}